\documentstyle{article}

\font\elevenbf=cmbx10 scaled\magstep 1
\font\elevenrm=cmr10 scaled\magstep 1
\font\elevenit=cmti10 scaled\magstep 1

\newcommand{\beq}{\begin{equation}}
\newcommand{\beqn}{\begin{eqnarray}}
\newcommand{\eeq}{\end{equation}}
\newcommand{\eeqn}{\end{eqnarray}}
\def\slash#1{\setbox0=\hbox{$#1$}#1\hskip-\wd0\hbox to\wd0{\hss\sl/\/\hss}}
\renewenvironment{thebibliography}[1]
 { \elevenrm
   \begin{list}{\arabic{enumi}.}
    {\usecounter{enumi} \setlength{\parsep}{0pt}
     \setlength{\itemsep}{3pt} \settowidth{\labelwidth}{#1.}
     \sloppy
    }}{\end{list}}

\begin{document}
\begin{center}{{\elevenbf Consistent Quantum Expansion Around
Soliton Solutions\\}
\vglue 1.0cm
{\elevenrm Pankaj Jain \\}
\vglue 0.7cm
\baselineskip=13pt
{\elevenit  Department of Physics and Astronomy \\}
\baselineskip=12pt
{\elevenit The University of Kansas\\}
\baselineskip=12pt
{\elevenit Lawrence, KS 66045-2151\\}
\vglue 2.0cm
{\elevenrm ABSTRACT}}
\end{center}
\vglue 0.3cm
{\rightskip=3pc
 \leftskip=3pc
 \elevenrm\baselineskip=12pt
 \noindent
I show that a standard application of the semiclassical techniques to
1+1 dimensional field theories, as originally discussed by Dashen,
Hasslacher and Neveu, explicitly violates the Poincare algebra.
This problem is traced to the incorrect regularization of the
ultraviolet divergences and can be resolved by using a
different regularization. I further show that in the case of the
doublet solutions in the sine-Gordon theory
 the semiclassical treatment given by Dashen,
Hasslacher and Neveu leads to ambiguous results which depend on the
choice of the renormalization counterterm. I discuss a
consistent weak coupling expansion which does not suffer from this
problem.}
\vfill
\noindent
To appear in Proceedings of DPF92, Meeting of the American
Physical Society (Fermilab 1992).
\eject
{\elevenbf\noindent 1. Introduction}
\vglue 0.1cm
\baselineskip=14pt
\elevenrm
The procedure of calculating quantum corrections to classical
soliton solutions was developed in the mid seventies [1,2]. The
formalism is based on the assumption that the quantum corrections
are small compared to the classical solution. The basic idea
can be explained by taking the simple example of 1+1 dimensional
sine-Gordon theory. This theory admits both time independent
and time dependent soliton solutions. For simplicity I consider first
the time independent classical solution $\phi_{cl}(x)$. The
quantum corrections can be calculated by expanding the field
$\phi$ as,
\beq
\phi(x,t) = \phi_{cl}(x-Z(t)) + \sum_n q_n(t)\psi_n(x-Z(t))
\eeq
Here I have introduced the collective coordinate Z and the fluctuation
coordinates $q_n$. The quantum hamiltonian can then be obtained
easily by introducing any set of orthonormal coordinates $Q_n$
and making a change of variables to the set $Z,q_n$.

The formalism described above has been applied to the 1+1
dimensional sine-Gordon as well as the $\phi^4$ field theory
by several authors. This application, however,
did not respect Poincare invariance or multiplicative renormalizability.
It turns out that these difficulties arise because of
incorrect regularization and because of the fact that certain
reasonable semi-classical techniques, valid for finite number of
degrees of freedom, fail in quantum field theory.

The fact the Poincare algebra is not satisfied can be easily checked
by evaluating the matrix elements of the commutator of momentum
with the boost generator between one soliton state. Explicit
calculation shows that this is not equal to the -i times the
energy of the soliton, if we follow the procedure described
 in Ref. [1]. The problem is traced to the incorrect choice
of regularization to regulate the ultraviolet divergent
sum over the frequencies. The regularization used in Ref. [1] is,
\beq
A_1 = \sum_{n=-N}^N \tilde\omega_n -  \sum_{n=-N}^N \omega_n
\eeq
where $\omega_n=\sqrt{k_n^2+m^2}$
 are the frequencies in the vacuum sector and
$\tilde\omega_n=\sqrt{q_n^2+m^2}$ are the frequencies in the soliton sector,
m being the meson mass.
The problem is easily corrected by choosing a different regulator
such as,
\beq
A_2 = \sum_{n=-\infty}^\infty \left[\tilde\omega_n f(t,q_n)
 -  \omega_n f(t,k_n)\right]
\eeq
where $f(t,q_n)$ is a regulating function such that it goes to 1 as
t goes to 0 and falls off sufficiently fast for large $|q_n|$ so
as to regulate the two sums.
The fact that some regulators fail to give reasonable answers is
somewhat disturbing but not too surprising. The same problem occurs
if we calculate the Casimir energy between two parallel plates
in QED. The result given by regulator in Eq. 2 is infinite and
has the wrong sign. The correct result is obtained by using the regulator in
Eq. 3.

We next examine the classical doublet solutions of the
sine-Gordon equation. The semi-classical
analysis of these states is rather interesting since the
quantum expansion used by Dashen, Hasslacher and Neveu (DHN)
in Ref. [1] seems to yield exact answers.
However as we have argued above, the regularization used in
Ref. [1] violates Poincare invariance and therefore it is somewhat
surprising that it yields exact results.
A careful analysis
of the calculation of quantum corrections to the doublet states,
however, reveals that the exact answer is obtained only by a
specific choice of the mass counterterm. A different choice of
the counterterm gives a wrong answer if the same
procedure is used as the one used in Ref. [1] for the doublet
solutions, thereby leading to a breakdown of multiplicative
renormalizability.

To discuss the problem in more detail I briefly review the
semi-classical technique applied to the sine-Gordon doublet states in
Ref. [1]. The authors consider G(E), defined as,
\beq
G(E) = Tr\left[{1\over E-H}\right]\; = \; i\, Tr\int dT\, exp[i(E-H)T]
\eeq
where $\phi_{\tau}(x,t=0)=\phi_{\tau}(x,t=T)$,
$\phi_{\tau}$ being in the classical
doublet solution.
By making an expansion around the classical doublet solution and keeping
only the leading order corrections to the classical solution DHN show that
\beq
G(E) = \int{dM\over 2\pi i}G^0(M,E)\bar G(M)
\eeq
\beq
\bar G(M) = -\left[{-i\over 2\pi}\right]^{1/2}\sum_l\int^\infty_0
\tau d\tau\sqrt l
\left[d\bar S/d\tau\over M\right]^{1/2}|d^2\bar S/d\tau^2|^{1/2}
exp[il(\bar S+M\tau)]
\eeq
where $\tau$ is the time period for one cycle. DHN perform the integration
over $\tau$ by using the stationary phase approximation which yields,
\beq
M = -{\partial \bar S\over \partial\tau}
\eeq
\beq
\bar S = S_{cl}(\phi_\tau) - {1\over 2}\sum_i\nu_i + S_{ct}(\phi_\tau)
\eeq
where $S_{ct}$ is the mass counterterm in the action.
The poles in G(E) are then found to be at,
\beq
W(M_N) =  S_{cl}(\phi_\tau) - {1\over 2}\sum_i\nu_i + S_{ct}(\phi_\tau)
+ E\tau(M) = 2N\pi
\eeq
where N=1,2,...$<8\pi/\gamma$, $\gamma$ is a function of the mass
and the coupling constant and is defined in [1].

However as shown in [3], the above procedure does not yield unique
results for $M_2/M_1$. The exact result was obtained in [1] by
a specific choice of mass counterterm. If we choose a different
counterterm and follow the same procedure as outlined above we get
a wrong answer [3].

This problem can be corrected if the integration over the time period
in equation 6 is performed by applying the stationary phase
approximation only to classical action. The quantum corrections are
then calculated by making a weak coupling expansion around this
stationary point. As shown in [3] the result obtained for $M_n/M_j$
is independent of the choice of counterterm to order $\lambda^3$,
$\lambda$ being the coupling constant, and
agree with the standard perturbation theory result to this order.
As expected, the result does not agree with the higher order two loop
result.

In conclusion we have discussed some difficulties with a standard
application of the semiclassical techniques to 1+1 dimensional field
theories. We have shown that although some regularization techniques violate
 Poincare algebra it is possible to choose one which does not
suffer from this problem. We have also argued that in order to get
unambiguous results it is necessary to make a weak coupling expansion
around the classical solution. The alternate procedure used in [1],
which seems to go beyond the weak coupling expansion, fails to
give unambiguous results in field theory.
\vglue 0.5cm
{\elevenbf \noindent 5. Acknowledgements \hfil}
\vglue 0.4cm
This work was
supported in part by the Department of Energy under grant No.
DE-FG02-85-ER40214.
\vglue 0.5cm
{\elevenbf\noindent 6. References \hfil}
\vglue 0.4cm

\end{document}